\documentclass[useAMS,usenatbib]{mn2e}

\usepackage{graphicx}

\title[Stars and dust in an ULIRG at $z = 1.135$]{
Stellar population and dust extinction in an ultraluminous infrared galaxy at $z = 1.135$}
\author[K. Kawara et al.]{K. Kawara$^{1}$\thanks{E-mail:
kkawara@ioa.s.u-tokyo.ac.jp}, S. Oyabu$^{2}$, Y. Matsuoka$^{3}$, 
Y. Yoshii$^{1}$, T. Minezaki$^{1}$, 
\newauthor
H. Sameshima$^{1}$, N. Asami$^{1}$,N. Ienaka$^{1}$, and T. Kozasa$^{4}$\\
$^{1}$Institute of Astronomy, the University of Tokyo, Osawa 2-21-1, Mitaka, 
Tokyo 181-0015, Japan\\
$^{2}$Institute of Space and Astronautical Science, Japan Aerospace 
Exploration Agency, \\
 3-1-1, Yoshinodai, Sagamihara, Kanagawa 229-8510, Japan\\
$^{3}$Graduate School of Science, Nagoya University, Furo-cho, Chikusa-ku, 
      Nagoya 464-8602, Japan\\
$^{4}$Department of Cosmosciences, Hokkaido University, Sapporo 060-0810, Japan
}
\begin{document}

\date{Accepted 2009 March 15}

\pagerange{\pageref{firstpage}--\pageref{lastpage}} \pubyear{2002}

\maketitle

\label{firstpage}

\begin{abstract}
 
We present the detailed optical to far-infrared observations of SST J1604+4304, an ULIRG at $z = 1.135$.
Analyzing the stellar absorption lines, namely, the CaII H \& K and Balmer H lines in the optical spectrum, 
we derive the upper limits of an age for the stellar population. Given this constraint, the minimum $\chi^2$ 
method is used to fit the stellar population models to the observed SED from 0.44 to 5.8\micron. We find 
the following properties.
The stellar population has an age 40 - 200 Myr with a metallicity 2.5 $Z_{\sun}$. The starlight is 
reddened by $E(B-V) = 0.8$. The reddening is caused by the foreground dust screen, indicating that dust 
is depleted in the starburst site and the starburst site is surrounded by a dust shell. 
The infrared (8-1000\micron) luminosity is $L_{ir} = 1.78 \pm 0.63 \times 
10^{12} L_{\sun}$. This is two times greater than that expected from the observed starlight, 
suggesting either that 1/2 of the starburst site is completely obscured at UV-optical wavelengths, 
or that 1/2 of 
$L_{ir}$ comes from AGN emission. The inferred dust mass is $2.0 \pm 1.0 \times 10^8 M_{\sun}$. This is 
sufficient to form a shell surrounding the galaxy with an optical depth $E(B-V) = 0.8$. 
From our best stellar population model - an instantaneous starburst with an age 40 Myr, 
we infer the rate of 19 supernovae(SNe) per year. Simply analytical models imply that 2.5 $Z_{\sun}$ in 
stars was reached when the gas mass reduced to 30\% of the galaxy mass. 
The gas metallcity is $4.8 Z_{\sun}$ at this point. The gas-to-dust mass ratio is then $120 \pm 73$.
The inferred dust production rate is $0.24 \pm 0.12 M_{\sun}$ per SN. If  1/2 of 
$L_{ir}$ comes from AGN emission, the rate is $0.48 \pm 0.24 M_{\sun}$ per SN.
We discuss 
the evolutionary link of SST J1604+4304 to other galaxy populations in terms of the stellar masses and the 
galactic winds, including optically selected low-luminosity Lyman $\alpha$-emitters and submillimeter 
selected high-luminosity galaxies.   

\end{abstract}

\begin{keywords}
cosmology: observations --- dust, extinction --- galaxies: individual: SST J1604+4304 --- galaxies: evolution 
--- galaxies: stellar content --- galaxies: starburst --- galaxies: high-redshift --- infrared: galaxies
\end{keywords}

\section{Introduction}

$IRAS$(Infrared Astronomical Satellite) observations discovered numerous infrared galaxies. 
At bolometric luminosities $> 10^{11} L_{\sun}$, infrared galaxies are the dominant 
population in the local Universe, being more numerous than optically selected 
starbursts, AGNs, and quasars \citep{sanders96}. By luminosity, infrared galaxies are classified into 
luminous infrared galaxies(LIRGs) with $L_{ir} > 10^{11} L_{\sun}$, 
ultraluminous infrared galaxies(ULIRGs) with $L_{ir}> 10^{12} L_{\sun}$, and 
hyperluminous infrared galaxies(HyLIRGs) with $L_{ir}> 10^{13} L_{\sun}$. The ratio of infrared 
to visible luminosity, $L_{ir}/L_B$ increases with $L_{ir}$ \citep{soifer89}. Although there is 
evidence that an optically buried AGN may exist in LIRGs, there is similarly strong evidence that 
enhanced star formation is ongoing (\citealt{armus95}; \citealt{sanders96}). \citet{heckman90} found 
that their optical spectroscopic data support the superwind model in which the kinetic energy 
provided by supernovae (SNe) and winds from massive stars in starburst site drives a large-scale 
outflow.

The {\it Infrared Space Observatory (ISO)} source counts in the mid- and far-infrared extended 
our knowledge of infrared galaxies up to $z \sim 1$ 
(\citealt{taniguchi97};\citealt{oliver97};\citealt{kawara98}; \citealt{puget99}; 
\citealt{elbaz99};\citealt{serjeant00}; 
\citealt{sato03}; \citealt{kawara04}),and revealed the number of ULIRGs rapidly grew with 
increasing redshift(\citealt{genzel00}; \citealt{chary01}; \citealt{heraudeau04}; \citealt{oyabu05}). 

The next large advance was 
made by the {\it Spitzer Space Telescope} with the sensitivity to probe infrared galaxies at $z = 2- 4$
(see review in \citealt{soifer08} reference therein). \citet{reddy06}, using deep {\it Spitzer} 24\micron\ 
observations, examined star formation and extinction in optically selected galaxies, near-infrared-selected 
galaxies, and submillimeter-selected galaxies (SMGs) at $z \sim 2$. {\it Spitzer} provided a powerful 
probe to measure redshifts of SMGs, the most luminous infrared galaxy population as HyLIRGs \citep{pope08a}.
Based on {\it Spitzer} mid-infrared data, the diagnostic diagram to separate starburst- and AGN-dominated 
infrared galaxies has been developed(e.g., \citealt{stern05}; \citealt{pope08b}; \citealt{polletta08}), 
and it is suggested that $L_{ir}$ in SMGs as well as 
ULIRGs at $z \sim 2$ is dominated by star formation\citep{pope08a}, while many SMGs contain an AGN as 
evidenced by their X-ray properties. \citet{dey08} uncovered a new population of dust-obscured galaxies 
(DOGs). These galaxies have $L_{ir}$ comparable to ULIRGs. However, their $L_{ir}/L_B$ is greater than 
local ULIRGs. DOGs and SMGs as well as high-redshift ($z \sim 1$) ULIRGs are generally 
considered to be young and massive galaxies, although the nature of the stellar populations is far from being 
understood. To understand the evolutionary link of these infrared galaxies to optically selected and 
near-infrared-selected galaxies, it is crucial to observe stellar absorption lines which indicate the age 
of the stellar populations. 
 
During the course of searching for distant objects in a cluster field, SST J1604+4304 
drew our attention with its red optical-3.6\micron\ color. Follow-up study revealed this object is 
an ULIRG at $z \sim 1.135$ and its energy source is young stars with an age $< 200$ Myr. 

We present the data consisting of optical spectroscopy and photometry from the optical to 
far-infrared in section 2. We derive an age and metallicity of the stellar population,
and show a foreground dust screen is plausible in Section 3. 
In Section 4, mid- and far-infrared emission by dust and energetics are discussed. 
In Section 5, we discuss the inferred size of a dust shell surrounding the galaxy, the SN rate and 
metallicity expected from the broad-band SED analysis, and the dust production rate per SN.
The evolutionary link of infrared galaxies to optically selected galaxies is discussed based on the  
galactic wind models of elliptical galaxies.

We adopt $H_0$ = 70 km s$^{-1}$ Mpc$^{-1}$, $\Omega_m = 0.3$, and
         $\Omega_{\Lambda} = 1 - \Omega_m$ throughout this paper. In this 
cosmology, the luminosity distance to the object is 7520 Mpc.
The flux density $F_{\nu}$ is used in units of $\mu$Jy, which is converted to AB
magnitudes through the relation  $m(AB) = -2.5 log(F_{\nu})$ + 23.9.

\section[]{Observations and data analysis}

This work is based on the archival data taken by the {\it Spitzer Space Telescope}, 
the {\it Hubble Space Telescope (HST)}, and the $Subaru$ Telescope and new photometric observations 
performed on the 
{\it United Kingdom Infrared Telescope(UKIRT)} and the $MAGNUM$\footnote{This telescope is
dedicated to the Multicolor Active Galactic Nuclei Monitoring (MAGNUM) project  
and is installed at the Haleakala Observatories in Hawaii \citep{yoshii02}.} telescope. 
In addition, the {\it Gemini Telescope North} was used to do optical spectroscopy.

\begin{figure}
\includegraphics[width=84mm]{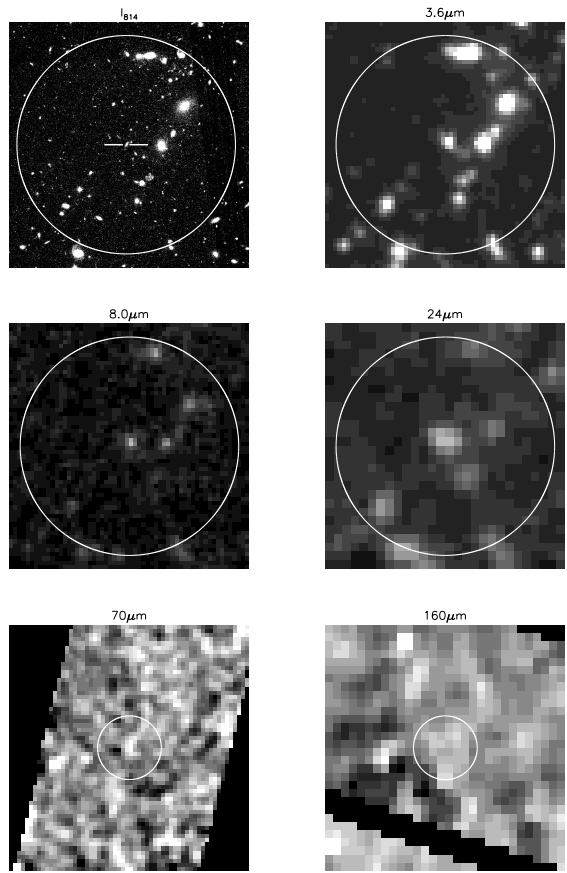}
\caption{Images centered at SST J1604+4304, where the size is 70\arcsec $\times$ 70\arcsec 
for $I_{814}$, 3.6\micron, 8.0\micron, and 24\micron, and 240\arcsec $\times$ 240\arcsec for 
70\micron, and 160\micron.
For all plots, north is at the top and east is to the left. We present 
32\arcsec radius circles which are the size of the aperture used for MIPS 160\micron\ photometry. 
Two bars in the $I_{814}$ image are used to indicate SST J1604+4304.
} 
\label{f_chart}
\end{figure}

\begin{figure}
\includegraphics[width=84mm]{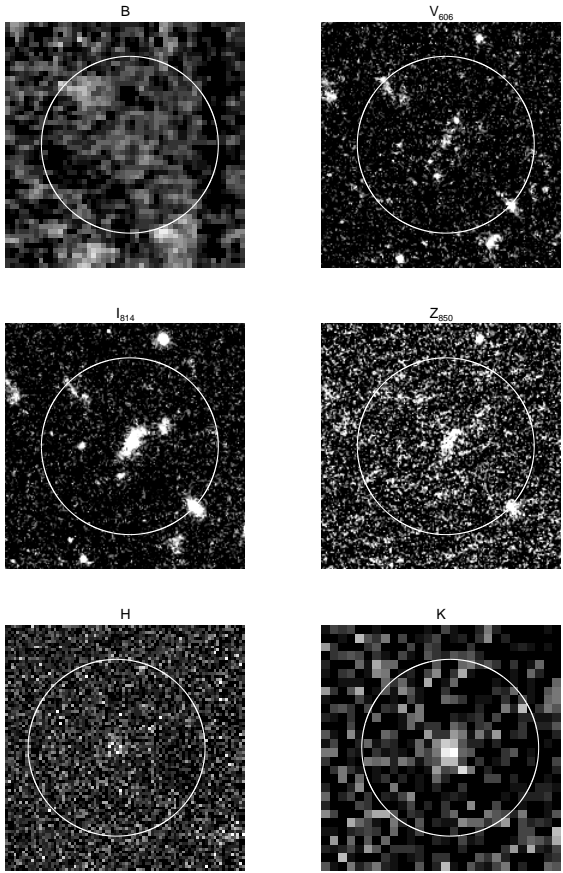}
\caption{10\arcsec $\times$ 10\arcsec images centered at SST J1604+4304. 
For all plots, north is at the top and east is to the left. We present 
3\farcs7 radius circles which are the size of the aperture used for IRAC 3.6-8.0\micron\ photometry. 
} 
\label{f_closeup}
\end{figure}

\subsection{Target}

SST J1604+4304 has been found at \\
\begin{flushleft}
R.A. 16$^h$ 04$^m$ 25\fs538\\
decl. +43\degr 04\arcmin 26\farcs55,\\
\end{flushleft}
\noindent
in the J2000 system in a field observed with the $Spitzer$ IRAC 
instrument. The optical counterpart in the $HST$ data is located within 
the 0\farcs4 offset accuracy requirements of $Spitzer$ \citep{werner04}.
Images of SST J1604+4304 are given in Figures \ref{f_chart} and \ref{f_closeup}.
The broad-band fluxes are given in Table \ref{t_fluxes}.

SST J1604+4304 is only 32\arcsec away from the center of the massive galaxy 
cluster CL 1604+4304 at $z = 0.90$ \citep{gal04}. Based on a weak-lensing 
analysis, \citet{margoniner05} find the mass contained within projected 
radius $R$ is $(3.69 \pm 1.47)[R/(500 kpc)] \times 10^{14} M_{\sun}$, 
corresponding to an inferred velocity dispersion 1004 $\pm$ 199 km s$^{-1}$.
Thus, using a singular isothermal sphere, we obtain the magnification $m = 1.17 \pm 0.09$ 
for SST J1604+4304 \citep{schneider92}.

\subsection{$Spitzer$ data}

All IRAC and MIPS data sets were retrieved from the $Spitzer$ public archive.
The IRAC data consist of four broadband images at 3.6, 4.5, 5.8,
and 8.0 \micron. The field-of-view is nearly 5\farcm2 $\times$ 5\farcm2
with a linear scale of approximately 1\farcs22 pixel$^{-1}$. 
The reader is referred to
\citet{fazio04} for a complete report on the in-flight performance of IRAC.
In the pipeline processing, ten BCD FITS images
were co-added into a single mosaic image per band with a technique similar
to 'drizzle'. The total effective exposure time is 1936 sec pixel$^{-1}$ 
per band. 
Photometry was performed using a 3 pixel (3\farcs7) radius aperture. 
As can be seen in the 3.6\micron\ image in Figure \ref{f_chart}, there are two sources 
which can contribute 
to the flux within the photometric aperture; a faint source is at 6\farcs5 to 
the southwest and a bright source 10\arcsec to the west. Using the point-spread 
function which is available on the IRAC website, their contributions were 
estimated to be 3.1 - 4.3 \% of the flux of SST J1604+4304 depending on the photometric 
band. The total fluxes after applying the aperture correction are given in 
Table \ref{t_fluxes}. The quoted errors include 5\% absolute calibration uncertainty 
that SSC\footnote{See Infrared Array Camera Data Handbook} recommends for general observers. 
Note that the color correction is very small and ignored.

The MIPS data were taken at 24, 70, and 160 \micron\ in 2004 March. In the pipeline 
processing, the mapped BCD data were rebinned and coadded to a single image with 
a linear pixel scale of 2\farcs45, 4\farcs0, and 4\farcs0 pixel$^{-1}$ 
for 24, 70, and 140 \micron, respectively. The total integration times are 93, 53, and 
84 sec pixel$^{-1}$. In the following analysis, we use the following 
post-BCD products, namely, mosaic images for 24\micron\ and filtered mosaic images for 
70 and 160\micron. 24\micron\ photometry was performed using a 7\arcsec radius 
aperture. The contributions, to the flux within the photometric aperture, from the 
two sources (see the 24\micron\ image in Figure \ref{f_chart}), located approximately 
at 15\arcsec to the west-northwest and 
13\arcsec to the southwest, were estimated using the point-spread function on the MIPS website. 
Their contributions are 9\% of the flux within the aperture. Then, the color correction 
for 500K blackbody and the aperture correction were applied. 

MIPS 70 and 160 \micron\ photometry 
was performed using 16\arcsec and 32\arcsec radius apertures, respectively. As seen in Figure \ref{f_chart}, 
SST J1604+4304 is faint in the far-infrared. We measured sky fluxes with the same 
aperture at 10 or 11 positions along the annulus with radii of 36\arcsec - 72\arcsec at 
70\micron\ and 96 - 150\arcsec at 160\micron. The flux averaged over 
these sky fluxes was subtracted from the flux measured within the object aperture 
centered at SST J1604+4304. The aperture correction and the color correction for 30K 
were applied. The detections are 2.5$\sigma$ and 5.0$\sigma$ at 70 and 160\micron, 
respectively. As seen in the 24\micron\ image in Figure \ref{f_chart}, there are 
several 24\micron\ sources within 
the 160\micron\ photometry aperture. Judging from their [3.6\micron]-[24\micron] color 
and 70 and 160 \micron\ brightness at their respective positions, the source at 13\farcs5 
to the west-northwest, the only source as red as SST J1604+4304, may contribute 1/3 of 
the flux within the aperture at most. The 160 \micron\ flux is 38.4 $\pm$ 7.65 mJy if the
contributions from this source can be ignored, while the flux is 25.6 $\pm$ 7.65 mJy if this 
source contributes 1/3 of the flux within the aperture. The real flux of SST J1604+4304 should 
be between them, and thus the flux is 32 $\pm$ 14 mJy. The aperture correction and the 
color correction for 30K blackbody were applied. Note that the 24\micron\ error in 
Table \ref{t_fluxes} includes 4\% absolute calibration uncertainty\footnote{Multiband 
Imaging Photometer for Spitzer (MIPS) Data Handbook}.

\subsection{Optical imaging data}

The $V_{606}$, $I_{814}$, and $Z_{850}$ image data were retrieved from the $HST$ public 
archive, where $V_{606}$, $I_{814}$, and $Z_{850}$ denote the $F606W$, $F814W$, and 
$F850LP$, respectively. The Advanced Camera for Surveys (ACS) Wide Field Camera (WFC) 
was used, which covers a field of 3\farcm4 $\times$ 3\farcm4 with a scale of 
0\farcs05 pixel$^{-1}$. Four image data were combined into a single co-added
image per band. The total exposure times are 4840 sec pixel$^{-1}$ for $V_{606}$ and 
$I_{814}$ and 6000 sec pixel$^{-1}$ for $Z_{850}$. The $I_{814}$ image in Figure 
\ref{f_closeup} shows that SST J1604+4304 has irregular morphology extending 1\farcs6 
in the NW direction with a 0\farcs6 width. To determine the total flux of this object, 
$I_{814}$ photometry was performed using 0\farcs8, 1\farcs0, 1\farcs2, 
and 2\farcs0 radius apertures. The $I_{814}$ flux grows from 0\farcs8 to 1\farcs2, while
it barely increase from 1\farcs2 to 2\farcs0. Note that contributions from the two 
nearest sources were subtracted from the flux measured with the 2\farcs0 radius 
aperture. Thus, the flux measured with the 1\farcs2 radius aperture represents the 
total flux of SST J1604+4304. The ratio of the flux with a 0\farcs8 aperture to a 
1\farcs2 aperture is 1.24 $\pm$ 0.05. $V_{606}$ and $Z_{850}$ photometry was performed 
using a 0\farcs8 aperture. The resultant fluxes were converted to the total fluxes 
by multiplying 1.24 $\pm$ 0.05, because we did not measure significant changes in 
optical colors across the target. The total fluxes are given in Table \ref{t_fluxes}.
The errors in Table \ref{t_fluxes} include the error of aperture conversion from 0\farcs8 
- 1\farcs2 and the correction for correlated noise in drizzling.

$Subaru$ $B, V, R_c, I_c, z$ data were retrieved from the SMOKA science achieve. 
The data were taken with Suprime-Cam (SUP) imager which has a field-of-view of 
34\arcmin $\times$ 27\arcmin with a scale of 0\farcs2 pixel$^{-1}$. 
The total exposure time is 2880, 2520, 3600, 1680, 3300 sec pixel$^{-1}$ 
after co-adding 4 to 11 image frames, using the standard SUP script \citep{yagi02}.
Flux scaling was performed observing 11 standard stars given by \citet{majewski94}. 
$I_c$ photometry were performed using three apertures with a radius of 
1\farcs0, 1\farcs2, and 2\farcs0, and it was confirmed the $Subaru$ $I_c$ brightness 
distribution is identical to that measured at the $HST$ $I_{814}$ photometry band. Thus, 
we use the 1\farcs2 radius aperture centered on SST J1604+4304 to obtain the fluxes 
given in Table \ref{t_fluxes}. The errors in Table \ref{t_fluxes} 
include 5\% photometric uncertainty.

\subsection{H/K imaging observations}

The $H$-band image was taken with the UIST (1-5 micron imager spectrometer) 
on $UKIRT$ in February 2006. UIST was used in the imaging mode with a field-of-view of
2\arcmin $\times$ 2\arcmin and a scale of 0\farcs12 pixel$^{-1}$. The total
exposure time is 1800 sec pixel$^{-1}$ after co-adding individual image data
with a integration time of 60 sec.
The sky condition was photometric with 0\farcs6 seeing.
The K-band image was taken with the multicolor imaging photometer (MIP)
on the 2 m $MAGNUM$ telescope
(\citealt{kobayashi98};\citealt{minezaki04}) in August 2006. MIP has a field-of-view of 1\farcm5
$\times$ 1\farcm5 with a linear scale of 0\farcs346 pixel$^{-1}$.
The total integration time is 2340 sec pixel$^{-1}$ after
co-adding individual image data. The weather condition was photometric with
1\arcsec seeing. Flux calibration was performed using 2MASS 16042631+4303413 
(H=14.36, K= 14.39) observed simultaneously with SST J1604+4304. Photometric 
uncertainty is 5\% and 9\% for H and K, respectively.

\begin{table}
 \centering
  \caption{Multiwavelength photometry of SST J1604+4304.}
  \begin{tabular}{crrrr}
  \hline
   Filter$^a$   &  Total flux        & $R_{ph}$$^b$  &  Instrument & UT \\
   band     &  $\mu$Jy              & arcsec     &             & yy/mm    \\
  \hline
$V_{606}$  & 0.224 $\pm$ 0.062 & 1.2      & $HST$ ACS   & 03/08 \\
$I_{814}$  & $1.05 \pm 0.08$   & 1.2      & $HST$ ACS   & 03/08 \\
$Z_{850}$  & $1.53 \pm 0.13$   & 1.2      & $HST$ ACS   & 07/02 \\
3.6\micron    & $60.35 \pm 3.16$  & 3.7      & $Spitzer$ IRAC & 04/03 \\
4.5\micron    & $53.76 \pm 2.79$  & 3.7      & $Spitzer$ IRAC & 04/03 \\
5.8\micron    & $55.11 \pm 4.07$  & 3.7      & $Spitzer$ IRAC & 04/03 \\
8.0\micron & $65.68 \pm 5.36$     & 3.7      & $Spitzer$ IRAC & 04/03 \\ 
24\micron  & $319 \pm 18.4$       & 7.0      & $Spitzer$ MIPS & 04/03 \\
70\micron  & $2150 \pm 870$       & 16      & $Spitzer$ MIPS & 04/03 \\
160\micron & $32000 \pm 14000$     & 32      & $Spitzer$ MIPS & 04/03 \\
\hline
$B$     & $0.0803 \pm 0.013$ & 1.2 & $Subaru$ SUP & 00/06 \\
$V$     & $0.111 \pm 0.015$  & 1.2 & $Subaru$ SUP & 05/05 \\
$R_c$   & $0.218 \pm 0.018$  & 1.2 & $Subaru$ SUP & 01/04,05 \\
$I_c$   & $0.689 \pm 0.042$  & 1.2 & $Subaru$ SUP & 01/04 \\
$z$     & $2.29 \pm 0.21$    & 1.2 & $Subaru$ SUP & 01/06 \\
$H$     & $5.40 \pm 1.19$    & 1.2 & $UKIRT$  UIST & 06/02 \\
$K$     & $20.06 \pm 3.35$   & 1.2 & $MAGNUM$ MIP  & 06/08 \\
\hline \hline
20cm    & $< 0.66$ mJy$^c$       & 2.7 & $VLA$$^e$           & 1995 \\
0.2-2 keV & $<2.1$ $^d$ & 15 & $XMM$$^f$ & 02/02 \\
2-10 keV & $<15$ $^d$ & 15 & $XMM$$^f$ & 02/02 \\
\hline
\end{tabular}
\medskip \\
\begin{flushleft}
$^a$ $V_{606}, I_{814}, \& Z_{850}$ denote F606W,F814W,\& F850LP, respectively.\\
$^b$ The radius of the photometric apertures which were used to obtain the total fluxes.\\
$^c$ To 3 $\sigma$.\\
$^d$ To 3 $\sigma$ in units of $10^{-15}$ erg cm$^{-2}$ s$^{-1}$.\\
$^e$ From the FIRST survey \citep{white97}.\\
$^f$ Observed by \citet{lubin04}. The FLIX online server was used to derive the upper limit.
\end{flushleft}
 \label{t_fluxes}
\end{table}

\subsection{Optical spectroscopy}

\begin{figure}
\includegraphics[width=84mm]{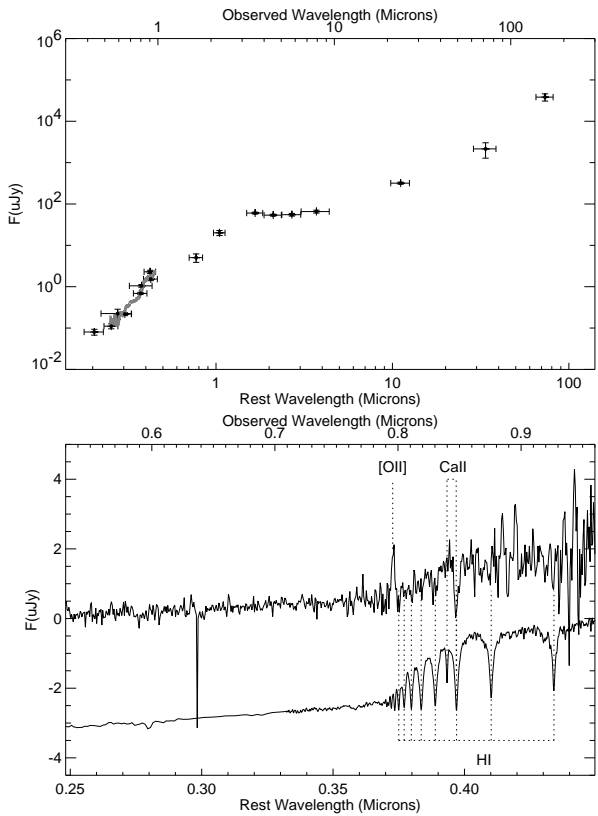}
\caption{The top panel shows UV to far-infrared SED for SST J1604+4304. 
The broad-band data are shown along with the median-smoothed optical 
spectrum ({\it gray}) taken with Gemini-North GMOS. The bottom panel shows the 
optical spectrum ({\it top}) together with the synthetic spectrum ({\it bottom}) 
at an age of 400 Myr, reddened by E(B-V) = 0.6} 
\label{f_sed}
\end{figure}

Optical spectroscopy was carried out using the Gemini Multi-Object Spectrograph (GMOS) 
on the Gemini-North telescope in 2007 August. The R150\_G5306 grating was used 
along with the OG151\_G036 filter, covering 5,000 - 10,000 \AA\ simultaneously with 
6.96 \AA/pixel after binned by 4 in the dispersion direction. 
The different wavelength centers (8,200 and 8,250 \AA) were set to fill 
the CCD gaps. A 1\farcs0 width slit was set so that the dispersion direction is 
perpendicular to the long axis of the target. Seven 1800 sec exposures were taken in the 
Nod \& Shuffle mode. Data were reduced in a standard manner using the IRAF 
GEMINI.GMOS package. The sky subtraction was made with the GNSSKYSUB routine. 
Wavelengths were scaled with CuAr spectra and the relative photometric calibration was performed using 
the standard star (HZ44). The results are shown in the bottom panel of Figure \ref{f_sed}.
The spectrum which is median-smoothed with a width of 40 pixles (i.e., $\sim$ 280 \AA), is compared 
with the photometry from the images in the top panel of Figure \ref{f_sed}. 
The two measurements agree with each other.

\section{stellar populations and dust extinction}

\subsection{Spectroscopic diagnosis}
 
The redshift of SST J1604+4304 is 1.135, as identified in Figure \ref{f_sed}.
The synthetic spectrum is also shown for comparison. 
This spectrum was made using the instantaneous-burst model \citep{bruzual03} at an age 400 Myr, 
reddened by E(B-V) = 0.6. 
The UV-optical spectrum is characterized by stellar absorption lines, which are 
H$_{\epsilon}$ at 3970 \AA, H$_{\zeta}$ at 3889 \AA, and probably higher order lines 
in the HI Balmer series as well as possible presence of 
the CaII HK absorption lines at 3968 and 3934 \AA. The emission line of [OII] at 3727 \AA\ 
is detected. No lines indicating the presence of an AGN, such as MgII$\lambda$2798 and 
[NeV]$\lambda$3346,3426, are detected. The 3 $\sigma$ upperlimit of the MgII EW (equivalent width) 
is 22 \AA, assuming a FWHM 6000 km s$^{-1}$ for the line width. This upper limit is small compared to 
52 \AA\ which is the MgII EW of the composite spectrum obtained from 184 quasi-stellar object 
by \citet{telfer02}. There are no signs indicating AGN emission at other wavelengths including X-ray and 
radio as shown Table \ref{t_fluxes}. Thus, we assume that the UV-optical spectrum is 
dominated by starlight.

The amplitude of the 4000 \AA\ discontinuity is known as a useful tool to measure 
the age of stellar populations(eg., \citealt{bruzual83}; \citealt{matsuoka08}). However,
it is not simple to correct for the reddening for heavily obscured objects like 
SST J1604+4304. Alternatively, the equivalent widths of CaII K and Balmer H lines are used 
to scale ages of star clusters (\citealt{rabin82}; \citealt{santos09}). 
Here We use the index which is defined as 
D(CaII) = 2 $\times$ EW(CaII K)/[EW(H$_{\zeta}$)+EW(CaII H + H$_{\epsilon}$)], where 
EW(CaII K) is the equivalent width for the CaII K line, EW(H$_{\zeta}$) for the H$_{\zeta}$
line, and EW(CaII H + H$_{\epsilon}$) for the blends of the CaII H \& H$_{\epsilon}$ lines.
The equivalent width is obtained by integrating across the line as 
EW = $\int_{\lambda 1}^{\lambda2 }(F^c_{\lambda} - F_{\lambda})d\lambda / \overline{F^c}$,
where $\overline{F^c} = \int_{\lambda 1}^{\lambda 2}F^c_{\lambda}d\lambda/(\lambda 2-\lambda 1)$.
$(\lambda 1, \lambda2)$ in \AA\ is (3871, 3909), (3926, 3940) and (3951, 3990) for H$_{\zeta}$, 
CaII K, and CaII H + H$_{\epsilon}$, respectively. $F_{\lambda}$ is the observed flux density and 
$F^c_{\lambda}$ is the continuum across the line, which is derived by fitting a linear function to 
assumed continuum points on either side of the line profile. Thus, we have D(CaII) = 0.03 $\pm$ 0.11 for 
SST J1604+4304. where EW(CaII K) = 0.4 $\pm$ 1.4 \AA, 
EW(H$_{\zeta}$) = 7.6 $\pm$ 2.1 \AA, and EW(CaII H + H$_{\epsilon}$) = 16.7 $\pm$ 2.2 \AA.

\begin{figure}
\includegraphics[width=84mm]{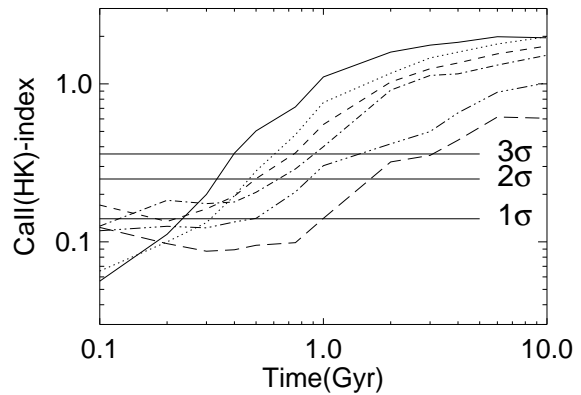}
\caption{The D(CaII) index as a function of ages of stellar populations. 
The index is defined as D(CaII) = 
2 $\times$ EW(CaII K)/EW(H$_{\zeta}$+CaII H + H$_{\epsilon}$). 
The metallicities are 0.005 $Z_{\sun}$ ({\it long dashes}), 0.02 $Z_{\sun}$ ({\it dash dot dot dot}),
0.2 $Z_{\sun}$ ({\it dash dot}), 0.4 $Z_{\sun}$ ({\it dashed}), 
1.0 $Z_{\sun}$ ({\it dotted}), and 2.5 $Z_{\sun}$ ({\it solid}). 
The three horizontal bar indicate the observed upper boundary of the D(CaII) index 
at 1, 2, \& 3 $\sigma$ upper limits.
}
\label{f_index}
\end{figure}

Figure \ref{f_index} shows D(CaII) measured in synthetic spectra for various metallicities as 
a function of ages of stellar populations. The synthetic spectra were obtained based on the 
instantaneous-burst models \citep{bruzual03} for metallicities $Z$ = 0.005, 0.02, 0.2, 0.4, 1, 
and 2.5 $Z_{\sun}$. The horizontal line shows the constraint from the observed 
D(CaII) at 1, 2, \& 3 $\sigma$ upper limits.
This diagram indicates that the inferred age of SST J1604+4304 depends on the metallicity;
the 3 $\sigma$ constraint is $<$1,000 Myr for $Z = 0.2 Z_{\sun}$, 
$<$750 Myr for 0.4 $Z_{\sun}$,$<$600 Myr for 1.0 $Z_{\sun}$,and $<$400 Myr for 2.5 $Z_{\sun}$.

The [OII]$\lambda$3727 line flux is $5.5 \pm 0.8 \times 10^{-17}$ ergs cm$^{-2}$ s$^{-1}$. 
Its implications for the star formation rate (SFR) will be discussed later.

\begin{figure}
\includegraphics[width=84mm]{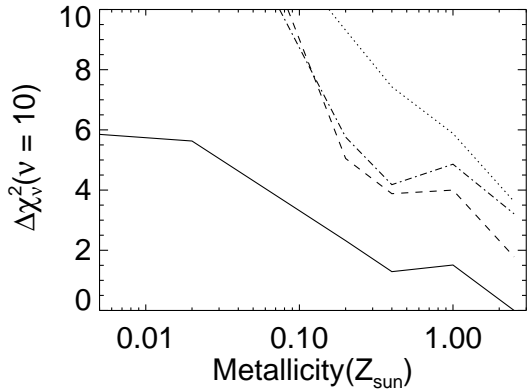}
\caption{$\Delta \chi_{\nu}^2$ versus metallicity, where
$\Delta \chi_{\nu}^2 = \chi_{\nu}^2 - \chi_{\nu}^2(min)$. $\chi_{\nu}^2(min)$, 
the smallest reduced $\chi^2 (=5.2)$, 
is obtained for $Z = 2.5 Z_{\odot}$ with the foreground Calzetti's dust screen. 
 The {\it solid} line shows 
$\Delta \chi_{\nu}^2$ for foreground dust screens with the Calzetti's extinction law,
the {\it dotted} line for foreground dust screens with the MW extinction law, 
the {\it dashed} line for internal dust models with the Calzetti's extinction 
law, and the {\it dash dotted} line for internal dust models with the MW 
extinction law. Note that $\chi_{\nu}^2$ was minimized for a given metallicity by changing 
ages $t$ and $E(B-V)$. }  
\label{f_zchi}
\end{figure}

\begin{figure}
\includegraphics[width=84mm]{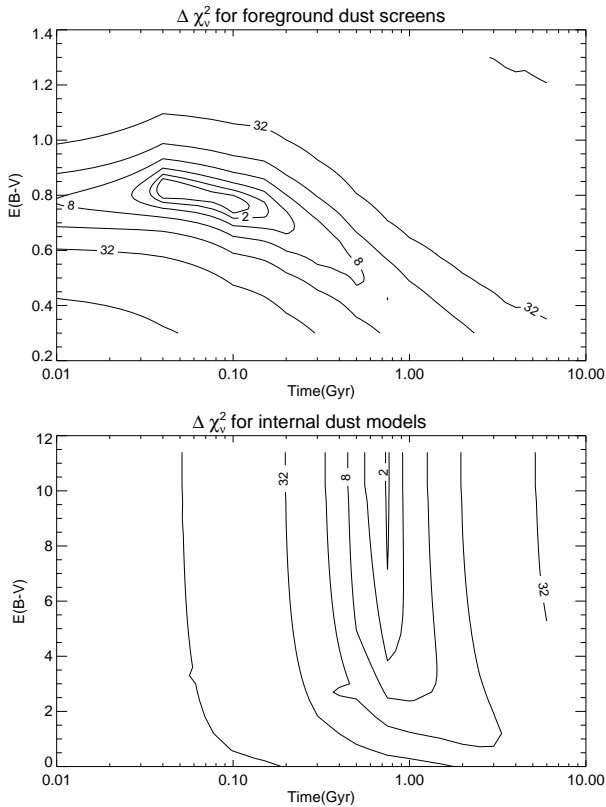}
\caption{$\Delta \chi^2_{\nu}$ maps for foreground dust screens ({\it top}) 
with $2.5Z_{\sun}$ instantaneous-burst models and 
internal dust models ({\it bottom}). The Calzetti's extinction law was used.
Notice that internal dust models
require much longer ages of stellar populations than those of uniform dust screens. } 
\label{f_contour}
\end{figure}

\subsection{Analysis of the UV to near-infrared SED}
 
We explore the nature of stellar populations and interstellar material in SST J1604+4304,
by fitting model spectra of stellar population models to the observed SED from the $B$ 
to 5.8\micron\ band. The 8.0\micron\ band is not included in this analysis, because 
emission from hot dust would be significant in this band.  
The evolutionary synthesis codes, BC03 \citep{bruzual03} are used to generate 
synthetic spectra of evolutionary stellar population models.  
The initial mass function (IMF) by \citet{chabrier03}(his table 1) is adopted 
with a lower and upper mass cutoff at 0.1 and
100 $M_{\sun}$, respectively. 

We use two extinction laws, namely, the starburst extinction curve by 
\citet{calzetti01} and the Milky way (MW) extinction curve by \citet{draine03}.
The starburst extinction curve was derived from the data in the UV and optical 
wavelength range of local starbursts and blue compact galaxies \citep{calzetti94}.
We consider two extreme cases of geometrical distribution of dust, namely, 
a {\it foreground dust screen} and {\it internal dust model}. In case of the 
foreground dust screen, dust is uniformly distributed in the screen which is physically distant from 
star clusters, so we observe 
$I^0 e^{-\tau}$ where $\tau$ is the optical depth of the dust screen and $I^0$ is  
unextincted radiation. In the internal dust model, dust and stars are uniformly mixed 
and the model schematizes the situation in which dust is purely internal to the staburst site,
so we observe $I^0 (1- e^{-\tau(1-\gamma)})/[\tau(1-\gamma)]$, where $\gamma$ is the albedo. 
Note that we assume the isotropic scattering.

\citet{calzetti01} gives the extinction curve in a narrow range from 0.12 - 2.2 \micron. 
To analyze multi-wavelemgth data from the UV to infrared, the extinction curve should be extended 
to wavelengths $ > $ 2.2 \micron.  The analytical expression given by \citet{calzetti01} 
cannot be used for this extension, because the expression makes extinction values sharply drop 
beyond 2.2 \micron. The Calzetti's extinction curve is almost identical to the \citet{draine03} 
MW curve from 0.6 -1.6 \micron.  We thus extend the Calzetti's curve to longer wavelenths
by adopting the MW extinction curve for wavelengths $ \ge$ 1.6 \micron.  
In addition, because \citet{calzetti01} does not provide the albedo, 
we adopt the albedo given in \citet{draine03} 
for the Calzetti's extinction law with removing the dip in the albedo which is attributed 
to the familiar 2175 \AA\ feature.

We start with instantaneous-burst models where all stars formed at once. 
The best model fit was searched for the observed SED by simultaneously optimizing 
the parameters, i.e., flux-scales, ages of stellar populations $t$, extinction $E(B-V)$ intrinsic to 
SST J1604+4304, and metallicities of the stellar populations($Z$). The ranges of the parameters 
are $t$ = 0.01 - 6 Gyr, $E(B-V)$ = 0.0 - 2.0 for the foreground dust screen and
$E(B-V)$ = 0.0 - 12.0 for the internal dust model, and $Z$ = 0.005 - 2.5 $Z_{\odot}$. 

The results are summarized in Figures \ref{f_zchi} and \ref{f_contour}.
The best-fit was obtained for $t$ = 40 Myr, E(B-V) = 0.8, and $Z = 2.5 Z_{\odot}$ 
with the foreground dust screen combined with the \citet{calzetti01} extinction law, 
resulting in a reduced chi square $\chi_{\nu}^2$ of 5.2. This $\chi_{\nu}^2$ is the minimum 
among all the models analyzed in this work. We thus refer to this minimum as
$\chi_{\nu}^2(min) = 5.2$, and the increment $\chi_{\nu}^2$ relative to $\chi_{\nu}^2(min)$ is 
represented as $\Delta \chi_{\nu}^2 = \chi_{\nu}^2 - \chi_{\nu}^2(min)$ \citep{avni76}. 
The approximate values of 
$\Delta \chi_{\nu}^2$ are 1.15, 1.83, and 2.77 for 68\%, 95\%, and  99.6\% confidence in this case(i.e., 
the degree of freedom is $\nu$ = 10).

\begin{figure}
\includegraphics[width=84mm]{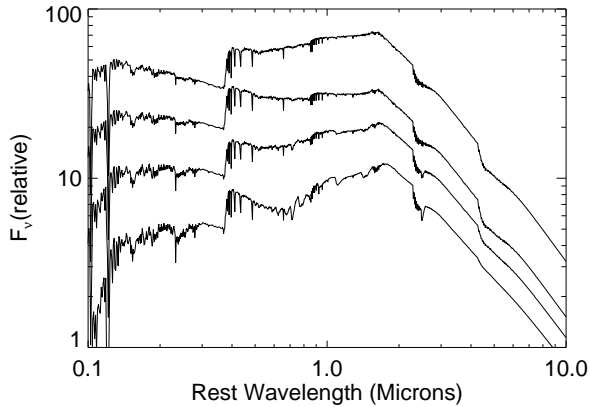}
\caption{Instantaneous-burst models with an age 40 Myr for various metallicities. From the top 
to the bottom, the synthetic spectra for $Z = 0.2, 0.4, 1.0, \& 2.5 Z_{\sun}$ are given. } 
\label{f_7000A}
\end{figure}

The foreground dust screen with the MW extinction law is ruled out, because of the too large 
$\Delta \chi_{\nu}^2 (>3)$. The internal dust models are ruled out either for two reasons:
(1)inferred ages of the stellar populations are too large to be fit within the range 
allowed by the D(CaII) index; (2) $\Delta\chi_{\nu}^2$ is too large, i.e., $\Delta\chi_{\nu}^2 > 3 $
for the $Z \le Z_{\sun}$ models and $\Delta\chi_{\nu}^2 = 1.8 $ for $Z = 2.5 Z_{\sun}$. 
For example, the best model for the internal dust geometry is obtained for 
$Z = 2.5 Z_{\odot}$. As can be seen in the bottom panel of Figure \ref{f_contour}, 
the inferred age is $t$ = 0.7 - 0.8 Gyr. This age is not allowed because
the age range must be $<$ 400 Myr, corresponding to the 3$\sigma$ D(CaII) limit for $Z = 2.5 Z_{\odot}$, 
as indicated in Figure \ref{f_index}. 
Figure \ref{f_zchi} shows that models with a rich metallicity give better fit to 
the data than those with a low metallicity. The sensitivity to metallicity arises 
from the broad concave feature around rest 7000\AA, as shown in Figure \ref{f_7000A}.
This feature is very weak in low metallicity spectra and becomes outstanding toward the high metallicity.

We also examined models with the exponentially declining star formation rate, which is 
represented by $\tau_e^{-1} exp(-t/\tau_e)$, where $\tau_e$ is the e-folding timescale and 
$t$ is the time after the onset of the first star formation. $\tau_e$ ranges from 0.01 - 
5.0 Gyr in a step of 0.5 dex and $t$ from 0.01 - 7 Gyr. The Calzetti's extinction law 
was used in the uniform dust screen geometry. All models with $Z \le Z_{\sun}$ 
are rejected because they have $\Delta \chi_{\nu}^2 > 2$. The D(CaII) index allows 2.5 $Z_{\sun}$ models 
to have $t \le$ 500 Myr, corresponding to 3$\sigma$ D(CaII) limit for 
exponentially declining star formation\footnote{For a given age of galaxies, D(CaII) is smaller 
in exponentially declining star formation than in instantaneous-burst, because 
in exponentially declining star formation stars are continuously forming.}. When $\tau$ is fixed and 
the other parameters are changed, the smallest values of $\chi_{\nu}^2$ are 
obtained for $E(B-V) \simeq 0.8$. In addition, $t$ correlates well with $\tau$. Using this correlation, 
we found $\tau \le $ 1000 Myr. Applying another constraint $\Delta\chi_{\nu}^2 < 1$, corresponding to the 
1 $\sigma$ deviation, we find that $\tau = $ 90 Myr and $t = $ 200 Myr are allowed for the exponentially 
declining star formation models.

The results are summarized in Table \ref{t_properties}. 
In Figure \ref{f_model}, we plot the best-fit(i.e., $\Delta \chi^2 = 0$)
instantaneous-burst model with $Z = 2.5 Z_{\sun}$ and an age of 40 Myr. 

\section{Dust emission and Energetics}

\subsection{Infrared luminosity}

\begin{figure}
\includegraphics[width=84mm]{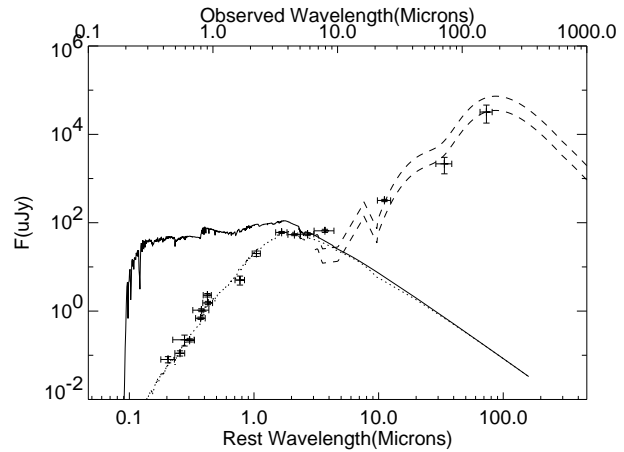}
\caption{UV to far-infrared SED for SST J1604+4304 along with the best-fit stellar 
population model with the Arp 220 template. The {\it solid} line 
shows the unreddened stellar spectrum and the {\it dotted} line the stellar 
spectrum with the uniform dust screen $E(B-V)=0.8$. This is the best-fit 
instantaneous-burst model for an age 40 Myr. The {\it dashed} lines represent the Arp 220 template 
fitted to three data points at 24, 70, \& 160 \micron\ ({\it upper}) 
and two data points at 70 \& 160 \micron\ ({\it lower}) , where the template is made based on 
\citet{carico90}, \citet{rigopoulou96}, and \citet{klaas97}.}
\label{f_model}
\end{figure}

As shown in Figure \ref{f_model}, dust emission is prominent at the 24, 70, and 160\micron\ 
bands. Because no data are available longward of 160\micron\, we fit the Arp 220 template to these 
data points to estimate the bolometric infrared luminosity. The template is made based on 
\citet{carico90}, \citet{rigopoulou96}, and \citet{klaas97}. In Figure \ref{f_model}, 
the upper dashed line represents the template which is fitted to the three data point 
at 24, 70 and 160 \micron, while 
the lower dashed line shows the template fitted to the two data points at 70 and 160 \micron.
The former seems to overestimate the luminosity, while the latter could underestimate the 
luminosity. We adopt the average of these two values for the real luminosity.
Then, the 8-1000\micron\ bolometric luminosity is  
$L_{ir} = 1.78 \pm 0.63 \times 10^{12} L_{\sun}$, classifying SST J1604+4304 as an ULIRG.
Note that $L_{ir}$ includes the correction for the flux magnification by 
cluster Cl 1604+4304, which is estimated to be 1.17.

\subsection{Energetics}  
The stellar luminosity, which is inferred from the best instantaneous-burst model, is 
$8.9 \times 10^{11} L_{\sun}$, accounting for only 1/2 of $L_{ir}$. If star formation activity dominates 
the bolometric output, the significant part of the starburst site is hidden by very opaque dust extinction which 
we cannot observe at UV to near-infrared wavelengths, implying the dust distribution in the foreground screen 
is clumpy. Otherwise, 1/2 of $L_{ir}$ is fueled by AGN emission.

The $XMM-Newton$ X-ray 0.2-10 keV and $VLA$ radio 20cm data listed in Table \ref{t_fluxes} do not support 
the AGN emission in SST J1604+4304. However, it should be noticed that 
the X-ray and radio observations are not deep enough to rule out a significant contribution from the AGN 
to the bolometric luminosity $L_{ir}$.
Mid-infrared data are widely used to search for obscured AGNs, because these wavelengths are much less 
affected by extinction and can penetrate through the dusty atmospheres surrounding the AGNs\citep{park08}.
Using the $f_{\nu}(8\mu m)/f_{\nu}(4.5\mu m)$ versus $f_{\nu}(24\mu m)/f_{\nu}(8\mu m)$ 
AGN diagnostic, \citet{ivison04} and \citet{pope08a} showed 
that infrared galaxies having $f_{\nu}(8\mu m)/f_{\nu}(4.5\mu m) < 1.9$ are 
starburst dominated in the mid-infrared.
SST J1604+4304 having $f_{\nu}(8\mu m)/f_{\nu}(4.5\mu m) = 1.22 \pm 0.12$ is 
thus starburst dominated. SST J1604+4304 has 
bluer mid-infrared colors than those of obscured AGNs and type1 QSOs studied by \citet{polletta08}.
\citet{stern05} showed a $[3.6\mu m]-[4.5\mu m]$ versus $ [5.8\mu m]-[8.0\mu m]$ 
color-color diagram for identifying AGNs.
SST J1604+4304 with $[3.6\mu m]-[4.5\mu m](Vega) = 0.35 \pm 0.08$ 
and $ [5.8\mu m]-[8.0\mu m](Vega) = 0.85 \pm 0.12$ lies
on the edge of the region which empirically separates AGNs from normal galaxies. 
\citet{yan07} performed mid-infrared spectroscopy of infrared luminous galaxies at $z \sim 2$ with the 
sample selection criteria $\nu f_{\nu}(24\mu m)/\nu f_{\nu}(8\mu m) \ge 3.16$. 
SST J1604+4304 having $\nu f_{\nu}(24\mu m)/\nu f_{\nu}(8\mu m) = 1.6$ does not meet the criteria of their 
sample. In conclusion, it is likely that mid- and far-infrared emission is starburst-dominated in 
SST J1604+4304, however we are unable to rule out the presence of AGN emission.  

\begin{figure}
\includegraphics[width=84mm]{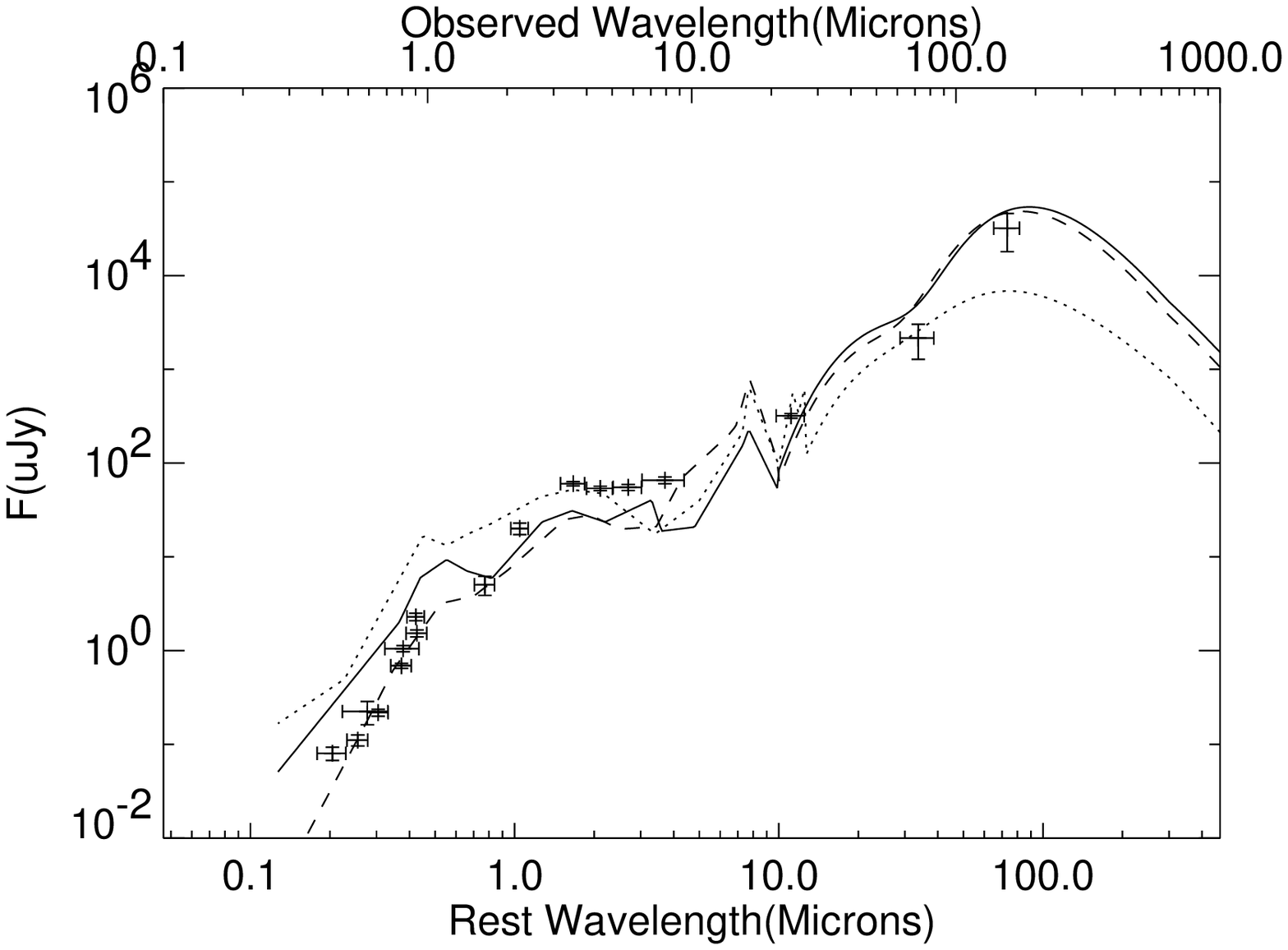}
\caption{UV to far-infrared SED for SST J1604+4304 compared with those of 
starburst-dominated galaxies, M 82({\it dot} local starburst), Arp 220({\it solid} local ULIRG), 
MIPS J142824.0+352619({\it dash} $z = 1.3$ HyLIRG). M82 data come from \citet{willner77}, \citet{dale07},
\citet{hughes94}, \citet{leeuw09}, and \citet{weedman04}, Arp 220 from \citet{carico90}, 
\citet{rigopoulou96}, and \citet{klaas97}, and MIPS J142824.0+352619 from \citet{desai06} and 
\citet{borys06}.
}
\label{f_ulirgs}
\end{figure}

We compare the UV to far-infrared SED of SST J1604+4304 with those of three starburst-dominated
galaxies, namely, M82(local starburst), Arp220(local ULIRG), and MIPS J142824.0+352619($z = 1.3$ HyLIRG)
in Figure \ref{f_ulirgs}. MIPS J142824.0+352619 with $L_{ir} = 3.2 \times 10^{13} L_{\sun}$ lacks 
any trace of AGN activity and has large PAH features and rich molecular gasses, 
indicating starburst-dominated mid-infrared emission
(\citealt{desai06};\citealt{borys06};\citealt{iono06}). 
As seen in Figure \ref{f_ulirgs}, there is a striking similarity 
between SST J1604+4304 and MIPS J142824.0+352619 in the overall SEDs from the UV to far-infrared. 

\subsection{Dust mass}
The dust mass $M_{dust}$ is related to the flux density $f_{\nu}$ as 
$M_{dust}/f_{\nu} = D^2/[\kappa_{\nu} B_{\nu}(T_{dust})]$ in the rest frame, where 
$D$ is the distance to the source and $\kappa_{\nu}$ is the absorption cross section per unit dust 
mass(e.g., \citealt{oyabu09}). When a modified blackbody spectrum $\nu^{\beta}B_{\nu}(T_{dust})$ is 
applied, $\beta$ and $T_{dust}$ should be used together as a set,
because $M_{dust}$ depends on $\kappa_{\nu} B_{\nu}(T_{dust})$, but not on $B_{\nu}(T_{dust})$ alone.
To derive $M_{dust}$, we will use the dust model for the Milky Way in the far-infrared 
\citep{draine03} in which $\kappa_{\nu} \propto \nu^2$ or $\beta = 2$. Assuming $\beta = 2$,
we obtain 32.5 K for the Arp 220 template and 35K for MIPS J142824.0+352619. This range of 
temperatures implies $M_{dust} = 2.0 \pm 1.0 \times 10^8 M_{\sun}$ for SST 1604+4304, where the uncertainty 
is dominated by the error in the 160 \micron\ flux.

\begin{table}
 \centering
  \caption{Pure star formation with no AGN for SST J1604+4304}
  \begin{tabular}{crr}
  \hline
   Quantity  & Instantaneous$^a$ & E-declining$^b$  \\
 \hline

$E(B-V)^c$                         & 0.83                        & 0.79  \\
$Z(Z_{\sun})^d$                    & 2.5                         & 2.5   \\
$\tau_e(Myr)^e$                    & --                          & 90     \\
$t(Myr)^f$                         & 40                          & 200  \\
$L/M_*^g$                 & 28                          & 14 \\
$L_{ir}(L_{\sun})$                 & $1.78 \pm 0.63 \times 10^{12}$  & $1.78 \pm 0.63 \times 10^{12}$ \\
$M_*(M_{\sun})$                    & $6.4 \pm 2.3 \times 10^{10}$         & $12.7 \pm 4.5 \times 10^{10}$ \\
$\Delta \chi_{\nu}^2$              & 0.0                         & 0.98  \\

\hline
\end{tabular}
\medskip 
\begin{flushleft}

$^a$ Instantaneous-burst models where stars form all at once.\\
$^b$ Exponentially declining star-formation models with SFR $\propto \tau_e^{-1}e^{-t/\tau_e}$, 
     where $t$ is the time after the onset of the initial star formation and $\tau_e$ is the 
     e-folding timescale.\\
$^c$ The \citet{calzetti01} extinction law is used in a foreground dust screen. \\
$^d$ Metallicity of the stellar populations. \\
$^e$ $e$-folding timescale for SFR. \\
$^f$ Time after the onset of the initial star formation.\\
$^g$ Luminosity to mass ratio in solar units, where $M_*$ is a stellar mass at $t$ = 0.\\
\end{flushleft}
\label{t_properties}
\end{table}

\section{Discussion}

For simplicity,  we temporarily assume that SST J1604+4304 is purely powered by star formation 
and at the end we will discuss the impact if SST J1604+4304 is AGN-dominated.

\subsection{Stellar mass}

The stellar mass $M_*$ can be estimated as $L_{ir}/(L/M_*)$, where $L/M_*$ is the luminosity to 
mass ratio for the stellar population models. Thus, we have $M_* = 6 - 13 \times 10^{10} M_{\sun}$ at 
$ t = 0$ as listed in Table \ref{t_properties}.
If this stellar mass was constantly built up during 40 - 200 Myr, the star formation 
rate is $SFR$ = 600 - 1650 M$_{\sun}$ yr$^{-1}$. Using the calibrations of SFR in terms of the 
[OII] luminosity (\citealt{kennicutt98}), after dereddening the [OII] flux, implies 13 - 500 
$M_{\sun}$ yr$^{-1}$. Considering that we are only looking at 1/2 of the starburst activity in the 
the optical, we obtain SFR = 26 - 1000 $M_{\sun}$ yr$^{-1}$, in good agreement with the 
previous estimate. 

\subsection{Comparison with other infrared galaxy populations}

SST J1604+4304 is a dusty, super metal-rich, young galaxy at $z = 1.135$ with $E_{B-V}$ = 0.8, 
metallicity 2.5 $Z_{\sun}$, and an age 40 - 200 Myr.
The age  is comparable to young, optical-selected populations like Lyman-$\alpha$ emitters (LAEs) and 
Lyman break galaxies (LBGs). Studying the $z \sim 3$ LAE sample, \citet{gawiser06} 
inferred ages of the order of 100 Myr with stellar masses $5 \times 10^8 M_{\sun}$ and 
no dust extinction. \citet{pirzkal07} showed $4 < z < 5.7$ LAEs had very young ages 
a few Mys with stellar masses $10^6 - 10^8 M_{\sun}$. \citet{sawicki98}, studying $z > 2$ 
LBGs, found that their LBGs were dominated by young stellar populations with ages
$<$ 200 Myr with stellar masses several $\times 10^9 M_{\sun}$ and moderate dust extinction
typically $E(B-V) \sim 0.3$. The $z = 2 - 3.5$ sample of LBGs by \citet{papovich01} has 
ages of 30 Myr to 1 Gyr with stellar masses $10^9 - 10^{11} M_{\sun}$ with 
$E(B-V)$ = 0.0 - 0.4. SST J1604+4304, having a stellar mass 6 - 13
$\times 10^{10} M_{\sun}$, is more massive and more dusty than these 
optically selected galaxies. 

SST J1604+4304 having $f_{\nu}(24\mu m)/f_{\nu}(R) > 1000$ meets the criteria for 24\micron-selected 
high-redshift ($z \sim 2$) dust-obscured galaxies (DOGs). 
DOGs and submillimeter-selected galaxies (SMGs) have extremely large $L_{ir}/L_B$
which cannot be accounted for by redshifting local ULIRGs (\citealt{dey08};\citealt{pope08b}).  
\citet{dey08} found nearly all of DOGs are ULIRGs with 
$L_{IR} > 10^{12} L_{\sun}$ and more than half of these have $L_{IR} > 10^{13} L_{\sun}$. 
The infrared luminosity of radio-identified SMGs ranges $L_{IR} = 2 \times 10^{11} - 10^{14} L_{\sun}$ 
\citep{chapman05}. A SFR 1000 $M_{\sun} yr^{-1}$, which corresponds to 
$L_{IR} = 5 \times 10^{12} L_{\sun}$ \citep{kennicutt98}, builds up a stellar mass 
of $10^{11} M_{\sun}$ in 100 Myr. Thus, it is considered that DOGs and SMGs are progenitors of 
present-day massive galaxies. If ages of stellar populations in DOGs and SMGs are derived, 
as we did for SST J1604+4304 by measuring the stellar absorption lines, 
the evolutionary link of DOGs and SMGs 
to normal ULIRGs and optically selected galaxies will be better understood. 

\subsection{SNe for dust and metal production}

In SST J1604+4304 with an age 40 - 200 Myr, most of the dust would be formed by type II supernovae (SNe II), 
because low-mass stars had not evolved to form dust in the expanding envelope. This may be the reason why 
our best model results in poor goodness of fit $\chi_{\nu}^2(min) = 5.2$. The Calzetti's extinction law 
may not be applicable to such young galaxies, because dust grains in local starbursts would be dominated by 
those formed in the outflows of low-mass stars and not by SNe II dust. 
In this regard, it is of great interest 
to use the SED of SST J1604+4304 to test extinction laws such as those calculated 
by Hirashita et al. (2005, 2008) that are predicted for dust formed in SNe II ejecta \citep{nozawa03}.

Our best model is the instantaneous-burst model with an age 40 Myr with a stellar mass 
$6.4 \pm 2.3 \times 10^{10} M_{\sun}$ at $t = 0$. The metallicity is 2.5 $Z_{\sun}$ and the dust mass is 
$M_{dust} = 2.0 \pm 1.0 \times 10^8 M_{\sun}$. Until 40 Myrs after the onset of the starburst, 
stars with a mass $> 8 M_{\sun}$ died with a SN explosion. The number of stars evolved into SNe in 40 Myr is 
$7.6 \times 10^8$ or 19 SNe per year. Adopting the nucleosynthesis models by \citet{nomoto06}, the total yield 
is 0.09 - 0.11 and the metal yield is 0.016 - 0.020 for the Chabrier IMF.\footnote{The original yields are 
given for the Salpeter IMF. These are converted to the Chabrier IMF by simply multiplying 1.64 to adjust the 
difference in the mass fraction of stars from 8 - 100 $M_{\sun}$ to those from 0.1 -100 $M_{\sun}$.} 
The gas and stetllar metallicities can be 
estimated applying simply analytical models based on the instantaneous recycling approximation. 
The gas metallicity 
is expressed as $Z_g(t)$ = $y(t)ln(1/f(t))$ where $f(t)$ is the ratio of the gas mass to the galaxy mass, and 
$y(t)$ is the ratio of the rate at which the metal is produced by the events of nucleosysnthesis and 
ejected into the interstellar gas to the rate at which hydrogen is removed from the interstellar gas by 
star formation \citep{searle72}. The stellar metallicity $Z_*$ is derived from 
$S(Z)/S_1 = (1 - f_1^{Z/Z_1})/(1 - f_1)$ and $Z_* = \int^{Z_1}_0 ZdS(Z)/ \int^{Z_1}_0 dS(Z)$,
where $S(Z)$ is the total mass of stars born up to a time when the metal abundance reached to the value $Z$, and 
$S_1, Z_1, f_1$ denote $S(t_1), Z(t_1), f(t_1)$, respectively \citep{pagel75}. 
In our case, $t_1$ is a time when the stellar metallicity reached
to $Z_* = 2.5 Z_{\sun}$. When the gas mass is reduced to $ f = 0.3$, we have $Z_* = 2.4 - 2.7 Z_{\sun}$ and 
$Z_g = 4.5 - 5.1 Z_{\sun} $. The inferred galaxy mass is $1.2 \times$ the stellar mass at $t =0$, 
because the total mass ejected from stars at the events of SN explosions into the interstellar gas is 0.1 
per unit stellar mass at $t =0$. The total gas 
mass is $2.3 \times 10^{10} M_{\sun}$. The gas-to-dust mass ratio is then $120 \pm 73$, which is comparable to 
the value obtained in a $z = 1.3$ HyLIRG ($198 \pm 53$) by \citet{iono06}.

The dust production rate is $0.24 \pm 0.12 M_{\sun}$ per SN. This is consistent with the numerical results of 
dust formation in SNe; for progenitor masses ranging 13 - 30 $M_{\sun}$, 2 - 5\% of the progenitor mass 
is locked into dust grains at SN explosions \citep{nozawa03}, 20 - 100\% of which is destroyed by processing 
through the collisions with the reverse shocks resulting from the interaction of SN ejecta and with the 
ambient medium \citep{nozawa07}.

\subsection{Foreground dust screen}

As demonstrated in Section 3.2, a foreground dust screen geometry is plausible for SST J1604+4304. 
This indicates that dust is depleted in the starburst site and the stellar 
spectra are reddened by the foreground dust screen which enclosed the starburst site. 
Let's suppose a spherical shell with a radius $l$ for the dust distribution, then we observe 
$f_{\nu} = 4 \tau_{\nu}B_{\nu}(T_{dust})\pi(l/D)^2$ in the rest-frame, 
where the thickness of the shell is $\tau_{\nu}$ which is 0.0074 at rest 80\micron\ corresponding 
to $E(B-V)$ = 0.83 with $R_V$ = 4.05 \citep{draine03}, 
$D$ is the distance from the observer to SST J1604+4304, and $T_{dust} = 32.5 - 35 K$. Converting 
this relation into the observer's frame, we obtain the radius $l$ = 4.5 - 5.5 kpc or 0.56 - 0.65\arcsec. 
This is comparable to the optical size of SST J1604+4304 which is approximately 
1.2\arcsec $\times$ 0.5\arcsec in the $I_{814}$ band. This supports the view that the galaxy 
is surrounded by the shell. Further support is the fact that there are no significant color gradients 
across the galaxy in the $HST$ images. 

\subsection{Evolutionary link to other galaxy populations}

As discussed by \citet{calzetti01}, starburst environments are rather inhospitable to dust; 
dust grains in the starburst site can be transported to a large distance in a relatively short time 
by radiation pressure(\citealt{ferrara91}; \citealt{venkatesan06}), as well dust grains that formed 
at SN explosions are processed and evolve in SN remnants \citep{nozawa07} - small size grains are 
quickly destroyed in SN remnants by sputtering. Thus, a cavity-shell structure 
is a natural geometry for the dust distribution in star-forming galaxies; the starburst site is inside the 
cavity where dust is depleted, and the opaque dust shell is surrounding the cavity. The shell is observed as
a foreground screen. Such foreground screens 
were found in local starburst galaxies (\citealt{calzetti94}; \citealt{meurer95}). 
In the foreground screen of SST J1604+4304, the dust would be unevenly distributed; we are only looking 
at part of the starburst site at UV and optical wavelengths through relatively transparent holes 
with $E(B-V)$ = 0.8, and the other part is completely obscured at these wavelengths.  
This explains that $L_{ir}$ is two times greater than the stellar luminosity derived 
from the broad-band SED analysis.

\citet{venkatesan06} and \citet{nozawa07} suggest that dust created in the first SN explosions can be 
driven through the interior of the SN remnants and accumulated in the SN shells, where second-generation 
stars may form in compressed cooling gases. Hence, galaxies may be observed as dust-free objects 
at the very beginning. When much dust formed in SN ejecta, the starburst site would become completely 
obscured by dust at UV to near-infrared wavelengths. This stage would correspond to extremely dusty 
objects like DOGs and SMGs. Then, the galactic winds
would turned on. \citet{heckman90} discussed that the galactic winds or superwinds frequently observed 
in starbursts and ULIRGs will sweep out 
any diffuse interstellar matter from the starburst site. 
Such galactic winds will turned on when the thermal energy of the 
gas heated by SN explosions exceeds the gravitationally binding energy of the gas. According to the models for
chemical evolution of elliptical galaxies by \citet{arimoto87}, galactic winds turn on later in more massive 
galaxies. The onset of a galactic wind is 350 Myr for a $10^{11} M_{\sun}$ galaxy, 
and the metallicity increases up to more than the solar value. These predictions seem to 
agree well with the age and the metallicity observed 
in SST J1604+4304, in which the galactic wind would just turn on and dust grains in the opaque shell
are pushed and moved outward, creating partially transparent holes in the shell.
This picture is consistent with the fact that dusty objects are deemed to be more abundant in massive 
galaxies than in less massive galaxies, although mid- and far-infrared observations are strongly biased 
to IR-luminous objects. This observational trend is 
expected, because galactic winds turn on later and thus the dust shell becomes more opaque 
for more massive galaxies,
resulting in longer time of the dust-obscured stage. This may be the reason why DOGs and SMGs are 
so infrared-luminous and massive.  

\subsection{Impact by an obscured AGN }

What would happen if the missing 1/2 of the bolometric luminosity comes from an AGN? In this case, 
the stellar mass ranging 0.1 - 8 $M_{\sun}$ is reduced to one half of that for the pure star formation.
The number of SNe is also reduced by the same amount, while the dust mass does not change. Thus, the dust 
production rate is increased by a factor of 2, i.e., $0.48 \pm 0.24 M_{\sun}$ per SN. 
This is within a range consistent with 
the model predictions(\citealt{nozawa03}; \citealt{nozawa07}). The metallicity is the same as 
derived for pure star formation, because the SN ejecta mass 
is also reduced by a factor of 2. If the AGN shines 
at the Eddington rate, the mass of the AGN is $2.5 \times 10^7 M_{\sun}$.

\section*{Acknowledgments}

We are grateful to Y. Tsuzuki for assisting the $UKIRT$ observation and to the anonymous referee for 
very useful comments.
This work has been supported in part by Grants-in-Aid for Scientific research
(17104002, 20340038) and Specially Promoted Research (20001003) from JPSP.

\end{document}